\documentstyle[preprint,aps,prb]{revtex}
\begin{document}

\title{Transmission Electron Study of Heteroepitaxial Growth in the
BiSrCaCuO System}

\author{A. Chaiken,  M.A. Wall, and R.H. Howell}
\address{Materials Science and Technology Division\\
Lawrence Livermore National Laboratory\\
Livermore, CA 94550}

\author{I. Bozovic, J. N. Eckstein, and G.F. Virshup}
\address{E. L. Ginzton Research Laboratory\\
Varian Associates, Inc.\\
Palo Alto, CA 94304-1025}
\date{\today}

\maketitle 

\begin{abstract}
Films of Bi$\rm _2$Sr$\rm _2$CaCu$\rm _2$O$\rm _8$ and Bi$\rm _2$Sr$\rm
_2$CuO$\rm _6$ have been grown using Atomic-Layer-by-Layer Molecular
Beam Epitaxy (ALL-MBE) on lattice-matched substrates.  These materials
have been combined with layers of closely-related metastable compounds
like Bi$\rm _2$Sr$\rm _2$Ca$\rm _7$Cu$\rm _8$O$\rm _{20}$ (2278) and
rare-earth-doped compounds like Bi$\rm _2$Sr$\rm _2$Dy$\rm _x$Ca$\rm
_{1-x}$Cu$\rm _2$O$\rm _8$ (Dy:2212) to form heterostructures with
unique superconducting properties, including superconductor/insulator
multilayers and tunnel junctions.  Transmission electron microscopy
(TEM) has been used to study the morphology and microstructure of
these heterostructures.  These TEM studies shed light on the physical
properties of the films, and give insight into the growth mode of
highly anisotropic solids like Bi$\rm _2$Sr$\rm _2$CaCu$\rm _2$O$\rm _8$.
\end{abstract}

\section{Introduction}

Bi$\rm _2$Sr$\rm _2$CaCu$\rm _2$O$\rm _8$, called 2212, is the
prototypical compound of a class of layered copper-oxide
superconductors which have been studied extensively due to their high
superconducting critical temperatures T$\rm _c$.  Other known compounds in
the BiSrCaCuO family, with the general formula Bi$\rm _2$Sr$\rm
_2$Ca$\rm _{n-1}$Cu$\rm _n$O$\rm _{2n+4}$, include Bi$\rm _2$Sr$\rm
_2$CuO$\rm _6$, called 2201, and Bi$\rm _2$Sr$\rm _2$Ca2Cu3O10, called
2223.  Like other cuprate superconductors, 2212 is a highly
anisotropic layered compound, with lattice parameters a = 3.818\AA and
c = 30.66\AA.\cite{torrance,sunshine} Despite the complexity of the
large unit cell, layer-by-layer growth of films has been achieved, as
has been demonstrated by the observation of oscillations in the
intensity of reflection high-energy electron diffraction (RHEED)
features during deposition.\cite{sakai} Many cubic materials, notably
GaAs, Si and various transition metals, can also be grown in a
layer-by-layer fashion, so in itself the occurrence of RHEED
oscillations is not remarkable.  The unusual aspect of the 2212
oscillations is that each molecular unit is composed of 14 layers so
that the cyclical growth pattern involves the formation of ordered
layers within the unit cell as well as the accumulation of completed
unit cells.  Putting 2212 films in a multilayer along with other
BiSrCaCuO compounds adds a further level of complexity.  Because of
the three different levels of ordering and the intrinsically strong
anisotropy, the dynamics of growth in a multilayer made up of 2212 and
its analogs can therefore be expected to be quite different from the
case of semiconductor or metallic multilayers.

Cross-sectional TEM is a useful tool for studying the microstructure
of multilayers and for classifying stacking faults in layered
structures.  Here TEM images have been used to study the morphology of
single films and heterostructures of BiSrCaCuO compounds which have
been grown using ALL-MBE.  This is a recently developed technique
involving sequential deposition of materials from metal vapor sources
(thermal effusion cells) in a highly reactive ozone
atmosphere.\cite{eckstein} The ALL-MBE technique has made possible
growth of high-quality multilayers and thereby fabrication of
Josephson junctions with novel tunnel
barriers.\cite{eckstein,eckstein2,bozovic} Described below are TEM
observations of a sampling of the heterostructures which can be grown
by ALL-MBE, including 2212/2201 multilayers and heterostructures with
2278 barrier layers.  In addition, the occurence of planar and line
defects in 2212 and 2201 will be described and discussed in relation
to the unusual growth modes of these materials.  The TEM micrographs
will be interpreted in comparison to image simulations produced by a
sophisticated electron-ray-tracing software package.

\section{Experimental}

The films used in this study were all grown using the ALL-MBE
technique, which has been described in detail
previously.\cite{eckstein2} ALL-MBE is a variant of molecular beam
epitaxy in which the composition and structure of a growing layered
film are controlled by sequential evaporation of the constituent
metals and their oxidation.  The growth is performed in a highly
reactive ozone atmosphere at elevated temperatures between 650 and
700$^{\circ}$C.  Growth is monitored in situ by observing reflection
high-energy electron diffraction patterns, and post-growth
characterization has been performed by x-ray diffraction, Rutherford
backscattering (RBS), Auger electron spectroscopy (AES), x-ray
fluorescence (XRF), secondary ion mass spectroscopy (SIMS), and atomic
force microscopy (AFM).  The RHEED observations confirm that abrupt
changes in crystalline phase can be made, namely that ALL-MBE allows
alternate layers of 2212 and 2201 to be grown as a
superlattice.\cite{eckstein} RBS, AES and EDAX measurements confirm
that the stoichiometry of the films is within a few percent of the
nominal value.\cite{howell} X-ray diffraction measurements show that
single-phase growth is possible using the ALL-MBE
technique.\cite{schlom} Total film thickness is typically on the order
of 1000 \AA.  The superconducting transitions of all films used for
TEM studies were measured resistively and found to be narrow, with a
transition temperature T$\rm _c$ similar to that reported
earlier.\cite{bozovic} Films like those used for TEM sample
preparation have been fabricated into tunnel junction
devices.\cite{eckstein2}

Preparation of high-temperature superconducting films for examination
by TEM has proven to be difficult.  The major difficulties are as
follows: differential thinning rates between substrate, film materials
and materials used to fabricate the cross-section specimen; the
inherent brittle nature of substrate and film materials; and the
production of specimen-preparation-induced artifacts such as ion
damage and amorphization.  The results of these problems are specimens
with only a small transparent region, fracture of the substrate and
delamination of the film.  To reduce and avoid the occurrence of these
specimen preparation difficulties, advanced techniques are employed.
The specimen preparation procedure described here employs a number of
steps to avoid mechanical damage and reduce artifacts.

The basic construction of the cross-sectioned specimen is similar to
the techniques describe by Newcomb\cite{newcomb} and
Bravman.\cite{bravman} The following is a detailed description of how
their procedure has been adapted for BiSrCaCuO films.

To begin with, a piece of blank Si wafer is epoxied to the SrTiO$\rm
_3$ substrate of the superconducting thin film.  The long direction of
the Si strip is oriented on the surface of the film-substrate
combination along the predetermined TEM viewing direction; i.e., [010]
or [110].  The substrate is then trimmed away with a low speed diamond
saw. The remaining Si/epoxy/film/substrate composite is laminated with
epoxy between two semi-circular brass rods and inserted into a brass
tube filled with epoxy.  After curing, discs measuring .5 mm thick are
sliced from the brass rod using a low-speed diamond saw.  Disc
specimens are lapped using a gravity feed holder from one side using a
succession of fine diamond lapping films to a thickness of .25 mm.
The specimen disc is then final polished on a vibrating polishing
machine for several hours.  Prior to lapping the second side, a Cu
grid is epoxied to the polished surface for support. The second side
is then lapped using the same diamond grit films to a final specimen
thickness of 100 $\mu$m (excluding the Cu grid).  The specimen is
dimpled from the second polished side directly on top of the
Si/SrTiO$\rm _3$ interface. The dimpling conditions are: 15 g load, 60
rpm, 2-4 $\mu$m diamond powders.  The final dimpled specimen thickness
prior to ion milling is approximately 15-20 $\mu$m.  Dimpling to
thinner dimensions often caused the substrate to fracture or the film
to delaminate.  Then a Gatan, Inc. PIPS ion mill is used to sputter
the film at a low angle of 4.5$^{\circ}$.  Low-angle ion milling in
combination with ion beam modulation is utilized to reduce the
difference in ion milling rates between the substrate, film, epoxy and
Si.  The first side of the specimen (dimpled side) is ion milled for
about 90 min.  The specimen is turned over and ion milled until
perforation.  After perforation both sides are again low-angle
ion-milled at 1.5 keV to reduce surface damage.  The specimen is
examined using low-resolution electron microscopy, and if necessary is
returned for further ion milling.

In order to characterize the layering and defect microstructure of
these atomically engineered films, high resolution electron microscopy
(HREM) was utilized.  A top entry JEOL 4000EX with a point-to-point
resolution of 0.17 nm was used to image atomic structure detail in
cross-sectioned specimens having film orientations of [010] and [110].
Experimental images were recorded near the scherzer defocus value
($\approx$ -45 nm) and the first cross-over of the contrast transfer
function ($\approx$ -75 nm).  Images were also recorded at other
defocus values at which specific layering and atomic structure detail
was best resolved.  The instrumental imaging conditions were as
follows: 400 keV accelerating voltage, C$\rm _s$ = 1.0 mm, focus
spread $\approx$8.0 nm, beam semi-divergence angle $\approx$0.8 mr and
an objective aperture radius of 0.096 nm$\rm ^{-1}$.

The atomic structure of the layering was confirmed by comparison of
the experimental images to those of simulated images.  Calculation of
the simulated images was performed using the MacTemPas software
package.\cite{mactempas} Published lattice constants were used as
inputs for 2201 and 2212, while lattice constants for 2278 and 1278
were estimated by using the same interatomic spacings as in the
equilibrium phases.  Since the specimen thickness and defocus
conditions are unknown for particular images, calculations were
performed for a range of values of these parameters.

\section{Results}

\subsection{2212 overgrowth on SrTiO$\rm _3$}

Figure~\ref{overgrowth} shows a high-resolution image of 2212
overgrowth on a SrTiO$\rm _3$ (001) substrate surface.  SrTiO$\rm _3$
has a lattice constant of 3.936 \AA, while that of 2212 in the c-axis
direction is 30.7\AA.\cite{torrance,sunshine} Examination of the image
shows that the ratio of the interplanar spacing in the SrTiO$\rm _3$
to that in the epilayer is in rough agreement with the ratio of the
lattice constants.  In addition, one can see good continuity between
planes of atoms in the SrTiO$\rm _3$ and in the overlayer, indicating
a coherent interface.

The most remarkable feature of the image shown in
Figure~\ref{overgrowth} is the lack of conformity of the 2212 film to
substrate surface features.  In the center of the image, there is an
asperity on the surface of the substrate.  The bright lines in the
2212 film are thought because of the image simulations (discussed
further below) to be BiO planes.  In the immediate vicinity of the
defect, there is a slight amount of disorder in the BiO planes, but
this disorder is healed on a very short lateral scale beyond which the
BiO planes are straight and unbroken.  In the vertical direction, the
BiO planes are continuous at only two 2212 molecular layers above the
substrate.  Somehow the 2212 layer is able to accomodate the
distortion caused by the substrate roughness within a single unit
cell.  The result is that after only two monolayers of 2212 growth,
the surface of the film is significantly smoother than the substrate
surface.  Apparently the energetics of the 2212 growth are such that
it is energetically costly to disrupt the bounding BiO planes.

The healing of defects at the 2212/SrTiO$\rm _3$ interface is in
contrast to the propagation of roughness which is often observed in
other epitaxial systems like Ge/Si\cite{headrick} and
GaAs/AlAs.\cite{sinha} In some cases the distribution of defects on
the surface of epitaxial films will be a replica of defects at the
film-substrate interface.  The shape of substrate-surface defects can
even be propagated through the entirety of thick multilayer
films.\cite{sinha} Propagative roughness is minimized in multilayers
meant for x-ray mirror applications by using an amorphous material as
one of the constituents.  For example, in Nb/Si and Mo/Si multilayers,
TEM studies have shown that the crystalline transition metal layers
have rough surfaces, but that the surfaces of the amorphous Si layer
above them are considerably flatter.\cite{chen} Presumably healing
occurs in the amorphous layer because the energetic cost of distorting
local bonding to accomodate interface roughness is less than in the
crystalline layers.  The surprising aspect of the 2212 overgrowth on
SrTiO$\rm _3$ is that the roughness of the substrate is healed within
a crystalline layer rather than an amorphous one.  The image suggests
that there is less of an energetic penalty associated with disrupting
the stacking of the BiO and CuO layers than with bending them.  This
assertion is consistent with the ability to grow BSSCO-family
materials with a varying number of CuO and Ca layers.

The strong tendency in the BiSrCaCuO materials to form continuous,
straight BiO planes likely contributes to the success of the ALL-MBE
technique with these materials. The naturally occurring layering in
the BiSrCaCuO superconductors may be favorable for techniques which
sequentially deposit the constituents, as ALL-MBE does.  Efforts to
grow the less anisotropic YBa$\rm _2$Cu$\rm _3$O$\rm _7$
superconductors using a similar technique have been less
successful.\cite{berkley}

The non-conformal growth of the 2212 may mean that coverage of
lithographically patterned steps in possible superconducting
microcircuits may be a problem.  On the other hand, the insensitivity
of the 2212 growth to the quality of the substrate surface undoubtedly
means that substrate preparation techniques need be less stringent in
this materials system than in many others.  For example, high-quality
overgrowth of Fe on Ag substrates requires painstaking surface
preparation because the vertical mismatch of the Fe and Ag lattice
constants causes disruption of the first three to four Fe monolayers
at atomic steps on the substrate.\cite{heinrich} This disturbance
occurs despite the good in-plane lattice match of Fe and Ag.  The
short range of the disruption of 2212 growth at asperities on the
SrTiO$\rm _3$ surface is further evidence of the inherently
anisotropic nature of the 2212 compound.

\subsection{2201/2212 Superlattices}

The precise and rapid changes in stoichiometry that are possible with
ALL-MBE allow growth of 2201/2212 superlattices with alternating
molecular units of the two constituents.  Each molecular unit is a
half-unit-cell of the full crystal structure.  A cross-sectional TEM
image of a 2201/2212 superlattice grown on SrTiO$\rm _3$ is shown in
Figure~\ref{superlattice}.  Superimposed on the micrograph are image
simulations for 2201 and 2212 that were produced with the same focus
and thickness parameters.  Qualitative agreement between the
calculated images and the micrograph is quite good.  Image simulations
with a variety of focus and thickness values convincingly demonstrate
that the bright lines bounded with dark regions are the BiO double
layers of 2212.

The high crystal quality of the 2201/2212 superlattice is not
surprising given the good lattice match between the two phases.  What
is striking in Figure~\ref{superlattice} is the layering of the 2201
and 2212, as is evident from the alternation of their respective
12.3\AA\ and 15.4 \AA\ molecular layer thicknesses.  X-ray diffraction
studies have previously given evidence for highly ordered growth of
2201/2223 multilayers with alternating half unit cells of the two
constituents.\cite{schlom} The low frequency of incomplete layers or
pinhole-type defects supports the previously published interpretation
of transport data on 2201/2212 superlattices, which showed that the
superconducting transition temperature T$\rm _c$ was not strongly
dependent on 2201 layer thickness.\cite{bozovic} The long lateral
continuity of the single molecular layers in this image confirms the
assertion that the 2212 layers in the superlattices are well isolated
from one another by the intervening layers of 2201.  The weak
dependence of the 2201/2212 multilayer T$\rm _c$ on the thickness of
the lower-T$\rm _c$ 2201 phase is therefore strong evidence of the
two-dimensional nature of superconductivity in 2212.\cite{bozovic} The
two-dimensional character of the superconductivity mirrors the
anisotropic nature of the crystal structure.

\subsection{2212/2278 Heterostructures}

A major advantage of the ALL-MBE method of film deposition is that its
precise control of layering makes possible the growth of BiSrCaCuO
phases which are not bulk equilibrium phases.  For example, ALL-MBE
has previously been used to grow films of the 2234 and 2245
compounds.\cite{schlom} The non-equilibrium BiSrCaCuO phases have the
bounding double BiO and single SrO layers that 2212 has, but the
number of CaCuO2 units internal to the unit cell may be varied in a
large range as long as the unstable layer is epitaxially stabilized by
growth on 2212.\cite{bozovic2} Figure~\ref{stick} shows the proposed
crystal structure of the 2278 phase and the well-known structure of
2212 for comparison.  The c-axis lattice constant of 2278 is
calculated to be 71.2 \AA\ by assuming that all interplanar distances
are the same as in 2212.  This assumption is probably not exact due to
the different charge-balance in 2278.  In addition to growing
non-equilibrium Bi$\rm _2$Sr$\rm _2$Ca$\rm _{n-1}$Cu$\rm _n$O$\rm
_{2n+4}$ phases, layers with only one bounding BiO layer have also
been deposited which have the stoichiometry Bi1Sr$\rm _2$Ca$\rm
_{n-1}$Cu$\rm _n$O$\rm _{2n+3}$.\cite{bozovic2} Thus a Bi-1278 layer
can be grown in a similar fashion to Bi-2278.

X-ray diffraction has been used to provide evidence for the successful
growth of several of these higher-n phases.\cite{schlom} Figure~\ref{2278}
shows a layer of the 2278 (n=7) material sandwiched between two thick
films of 2212.  While examination of TEM images is an inexact method
of lattice constant determination, the figure shows that the layer
spacing of the 2212 layers (15.4 \AA) is slightly less than half the
thickness of the nominal 2278 layer, which is calculated to be 35.6
\AA\ thick.  The 2278 phase is of considerable interest because when
doped with Dy it is useful as a tunnel barrier in Josephson
junctions.\cite{eckstein} 2278 and 1278 are also superconducting in their own
right.\cite{bozovic2,bozovic3}

TEM cross-sectional images (not shown) have also been taken of nominal
1278 tunnel barriers.  The nominal layer thickness of 1278 (which
should be a simple tetragonal rather than a body-centered tetragonal
structure) is calculated to be 32.5 \AA.  It was not possible from TEM
to distinguish them from the 2278 barriers and to determine whether a
single BiO layer structure has actually been achieved.  A microscopy
technique which provides elemental contrast\cite{pennycook} may be
necessary to get at this information or to determine the
rare-earth-atom spatial distribution in the Dy-doped barrier layers.
At this stage it is possible to report only that the transport
properties of tunnel junctions with 2278 and 1278 barriers are quite
different.\cite{bozovic2}

\subsection{Defects in 2212-based Heterostructures}

The crystal quality of 2201 and 2212 films grown on a SrTiO$\rm _3$
substrate by ALL-MBE is very high, as described above.  The only
defects which are commonly observed in these films are twin
boundaries, one of which is shown in Figure~\ref{twin}.  Part of the
image has the 2212 a-axis oriented normal to the plane of the paper;
this portion appears to have long, unbroken BiO planes.  The other
part of the image has the 2212 b-axis oriented normal.  The
b-axis-oriented portion of the image displays the incommensurate
modulation which has previously been observed in 2212
films.\cite{bando} It has been shown that the growth of untwinned
films via ALL-MBE is possible by using specially miscut
substrates.\cite{eckstein3} However, the special precautions necessary
to prevent twinning are not deemed to be worthwhile for routine
growths since the presence of twins is not believed to affect the
performance of Josephson junctions.

Occasionally more troublesome defects may occur.  These defects
include stacking faults and antiphase boundaries, as
Figure~\ref{stack} illustrates.  The stacking fault defects are most
easily visualized by observing the presence of the wrong number of BiO
planes, although presumably the SrO planes and CaCuO$\rm _2$ planes
may also be affected.  There are several places in the image of
Figure~\ref{stack} where apparently more than two BiO planes occur
together.  Nearby in the image, an arrow points to the spot where
layer spacings indicate that two phases of different stoichiometry
have grown together.  Estimates of the layer thicknesses for these
phases indicate that they may correspond to 2223, which has an extra
CaCuO${\rm _2}$ slab, and 3312, which has an extra BiO-SrO
unit.\cite{howell} Exactly why these off-stoichiometric phases and
stacking fault defects occur has not yet been determined.  For reasons
which are not well-understood, the frequency of occurrence of these
defects is greater above a 2278 or 1278 barrier than in a 2212 film
grown on a 2201 buffer layer.  It may be hypothesized that the
unfavorable charge balance in the unstable barrrier
layer\cite{bozovic2} may cause havoc in subsequent overgrowths.

RBS, EDAX and SIMS measurements indicate that the overall composition
of the films is within a few percent of the nominal
stoichiometry. However, defects which are visible on the surface of
the superconducting films with optical microscopy are found by RBS,
SIMS and EDAX to have a substantially different
stoichiometry.\cite{howell} These surface defects typically occur in
the corners of the films, where the deviation from the nominal
stoichiometry would be expected to be greatest because the thermal
effusion sources are not exactly on-axis.  A correlation between the
occurrence of these defects and poor Josephson junction performance
has been observed.\cite{howell} Whether any of the defects which are
visible in plan view with optical microscopy are identifiable with the
defects seen in TEM cross-sections is not yet clear.  The length scale
of the the stacking faults and antiphase boundaries which are observed
in cross-section is angstroms, while the size of optically observable
surface defects is microns.  Nonetheless, it may be that the nanoscale
imperfections in film structure serve as nucleation points for the
larger defects which are observed in plan view.

Despite the presence of planar defects and twins in the BiSrCaCuO
films, the overall impression that TEM observations leave is one of a
high degree of ordering.  The low-magnification TEM image shown in
Figure~\ref{lowmag} illustrates this point.  The bright layers in the
image are three barrier layers, each designed to consist of a single
molecular layer of 2278.  Clearly these barriers are continuous over
very large in-plane lengths, perhaps 1000 \AA.  The in-plane continuity
of the barriers is clearly larger than that out-of-plane, once again
illustrating the inherent anisotropy of BiSrCaCuO growth.  An estimate
of the average film non-stoichiometry can be made from an enlarged
version of the image in Figure~\ref{lowmag} by comparing the observed
total thickness of the film to the thickness of a perfect film with
the same layering sequence.  The comparison shows that the Bi:Cu ratio
of the film deviates from the target composition by only about 1\%.
The thickness is determined by the Bi:Cu ratio since it is this ratio
that determines which of the BiSrCaCuO compounds is formed.

\section{Conclusions}

Transmission electron microscopy is uniquely powerful when it comes to
imaging point or line defects in a solid, or when one wants to image
the structure of a single molecular layer in a film.  Thus, TEM
provides information about buried, localized features of thin films
which is difficult to obtain in any other fashion.

The TEM images taken as part of this study have strikingly confirmed
previous hypotheses about the structure of heterostructures and
multilayers grown with the ALL-MBE technique.  In particular, the TEM
images demonstrate that single layers of 2212 and 2201 can be grown as
part of a multilayer.  Furthermore, single layers of metastable or
unstable phases such as 2278 and 1278 can also be grown.  These TEM
micrographs provide the first direct evidence for the existence of
these phases. While growth flaws like twins, stacking faults and
antiphase boundaries do occur, especially in overgrowths on metastable
barrier layers, the images are most notable for the long, unbroken
continuity of most 2212 planes.  Up to this point, ALL-MBE growth of
BiSrCaCuO superconductors have not reached the perfection that is
observed in MBE growth of GaAs, AlAs and related compounds.  However,
planned improvements in the ALL-MBE process such as rotation of the
substrate during growth may make comparably low defect densities
possible.

The TEM images presented here support previous ideas about the
two-dimensional nature of superconductivity in the 2212
compound.\cite{bozovic} In addition, they provide new information
about the ways in which growth of inherently highly anisotropic phases
may differ from growth of cubic materials.  Studies of the growth of
high-temperature superconductors may in the end have as much to teach
us about anisotropic solids as about the fundamental nature of
superconductivity.

\paragraph*{Acknowledgements}

We thank Sandia National Laboratory and George Thomas for use of the
TEM.  We thank P.A. Sterne and M.J. Fluss for useful discussions,
R.G. Musket for the RBS measurements, J.M. Yoshiyama for the EDAX
measurements, and D.L. Phinney for the SIMS.  Part of this work was
performed under the auspices of the U.S. Department of Energy by
Lawrence Livermore National Laboratory under contract number W-7405-
ENG-48.  The work at Varian was supported in part by NRL and ONR via
contracts N00014-93-C-2055 and N00014-94-C-2011.

\clearpage

\begin{figure}
\caption{A high-resolution cross-sectional TEM image showing overgrowth
of a 2212 superconductor film on a SrTiO$\rm _3$ substrate (001) surface.
The view is along the (100) axis of the SrTiO$\rm _3$ substrate. Registry
between planes in the substrate and overlayer is nearly perfect,
resulting in a coherent interface.  The BiO planes are continuous
throughout the image, even directly above rough areas on the SrTiO$\rm _3$
surface.}
\label{overgrowth}
\end{figure}

\begin{figure}
\caption{A high-resolution cross-sectional TEM image of a 2201/2212
superlattice.  The view is along the (001) axis of the SrTiO$\rm _3$
substrate.  The alternation of the 15.4 \AA\ period of 2212 and the
12.3 \AA\ period of 2201 is readily identifiable.  Also shown are
image simulations of 2212 and 2201 which were calculated using the
known lattice constants of the materials.  Parameters for the
simulations are focus = -450, thickness = 200\AA.}
\label{superlattice}
\end{figure}

\begin{figure}
\caption{Hypothetical crystal structure of the Bi-2278.  The c-axis
lattice constant is 71.2 \AA, so the thickness of a single layer
should be 35.6 \AA.  The in-plane lattice constants are assumed to be
the same as in 2212.  The view is along the (001) axis.  Shown for
comparison is the well-known crystal structure of the 2212 phase.}
\label{stick}
\end{figure}

\begin{figure}
\caption{High-resolution cross-sectional image of a 2278 barrier layer
sandwiched between two thick films of 2212.  The view is along the
(001) axis of the SrTiO$\rm _3$ substrate.}
\label{2278}
\end{figure}

\begin{figure}
\caption{High-resolution cross-sectional image of a twinned section of
a 2212 film.  The layers on the right-hand side of the image have the
b-axis normal to the plane of the figure and show incommensurate
modulation.  Layers on the left-hand side have the a-axis normal to
the plane of the figure.}
\label{twin}
\end{figure}

\begin{figure}
\caption{Stacking defects in a 2212 film grown on SrTiO$\rm _3$ substrate.
View is along the (001) axis of the SrTiO$\rm _3$ substrate.  The antiphase
boundary is indicated by a black arrow.}
\label{stack}
\end{figure}

\begin{figure}
\caption{Low-resolution micrograph of a BiSrCaCuO film with three
barriers each consisting of a single molecular layer of 2278,
separated by thicker layers of 2212.  This image demonstrates the long
lateral coherence length which is typically observed in
BiSrCaCuO-based heterostructures grown by ALL-MBE.}
\label{lowmag}
\end{figure}

\end{document}